\documentclass[useAMS,usenatbib,usegraphicx]{mn2e}

\title[Young GCs in an old S0: Formation History of
NGC4570]{Young Globular Clusters in an old S0: \\ Clues to the Formation
  History of NGC~4570\thanks{Based on observations made with ESO Telescopes at
    the La Silla Observatory under programme ID 079.B-0511}}
\author[R. Kotulla, U. Fritze and P. Anders]{R. Kotulla$^{1}$\thanks{E-mail:
r.kotulla@herts.ac.uk}, U. Fritze$^{1}$ and P. Anders$^{2}$\\
$^{1}$Centre for Astrophysics Research, University of Hertfordshire, College
Lane, Hatfield AL10 9AB, United Kingdom\\
$^{2}$Sterrenkundig Instituut, Universiteit Utrecht, P.O. Box 80000,
NL-3508 TA Utrecht, The Netherlands}
\begin{document}

\date{Accepted April 05, 2008. Received March 26, 2008; in original form
  February 04, 2008}

\pagerange{\pageref{firstpage}--\pageref{lastpage}} \pubyear{2008}

\maketitle

\label{firstpage}

\begin{abstract}
  We here present our first attempt to use Globular Clusters as tracers of
  their parent galaxy's formation history. Globular Cluster Systems of most
  early-type galaxies feature two peaks in their optical colour distributions.
  Blue-peak Globular Clusters are generally believed to be old and metal-poor.
  The ages, metallicities, and the origin of the red-peak Globular Clusters
  are being debated. We here present our analysis of the ages and
  metallicities of the red peak Globular Clusters in the Virgo S0 NGC 4570
  using deep $K_s$-band photometry from NTT/SOFI (ESO program ID 079.B-0511)
  for the red-peak Globular Clusters in combination with HST-ACS archival data
  to break the age-metallicity degeneracy.

  We analyze the combined $g$, $z$, and $K_s$ spectral energy distribution by
  comparison with a large grid of GALEV evolutionary synthesis models for star
  clusters with different ages and metallicities. This analysis reveals a
  substantial population of intermediate-age (1--3 Gyr) and metal-rich
  ($\approx$ solar metallicity) Globular Clusters. We discuss their age and
  metallicity distributions together with information on the parent galaxy
  from the literature to gain insight into the formation history of this
  galaxy.

  Our results prove the power of this approach to reveal the (violent) star
  formation and chemical enrichment histories of galaxies on the basis of
  combined optical and near-infrared photometry.

\end{abstract}

\begin{keywords}
  galaxies: individual: NGC 4570 -- galaxies: star clusters -- galaxies:
  lenticular -- galaxies: formation.
\end{keywords}

\section{Introduction}
Globular Cluster (GC) systems are now recognized as powerful tracers of
their parent galaxy's formation history \citep{west04,fritze04,brodie06}.
From their age and metallicity distributions one can reconstruct the parent
galaxy's (violent) star formation and chemical enrichment histories all the
way from the very onset of star formation in the Early Universe to the
present.

Most early-type galaxies show bimodal color distributions for their GC systems
\citep[e.g.][]{gebhardt99,kundu01a,kundu01b,peng06}: A universal blue peak and
a red peak for which the color and height relative to the blue peak vary from
galaxy to galaxy. The blue peak GCs are generally agreed to be old and
metal-poor, the properties and origin of the red peak GCs is still unclear.
Scenarios for the formation of the red peak GCs range from {\em in situ}
formation of a secondary more metal-rich population of GCs within their parent
galaxy shortly after the first one \citep{Forbes+97} to major gas-rich mergers
\citep[e.g.][]{AshmanZepf92} and hierarchical accretion events involving
enough gas to trigger the formation of new GC populations \citep[e.g.][]{Beasley02}. The age distributions of the secondary GCs predicted by
these different scenarios are different: almost as old as the old and
metal-poor blue peak GCs but more metal-rich than those in the first case, of
some intermediate age reflecting the time of the gas-rich merger in the second
case, and broad or multi-peaked for a series of hierarchical accretion events
involving gas in the third case. The metallicities of the secondary, and
eventually any further generations of GCs should reflect the ISM abundances in
the merging or accreted objects at the time of merging or accretion. They
could, at most, be somewhat higher for those GCs that formed late enough in a
burst to already incorporate some enrichment during the burst itself.
Hierarchical accretion without gas and the formation of new generations of GCs
cannot explain the red-peak GCs since dwarf galaxies known so far contain old
and metal-poor, hence blue GCs.

To determine the ages and metallicities of the red-peak GCs in one of those
early-type galaxies with clear bimodality in its optical GC colour
distribution is the aim of our present investigation and should help constrain
the formation scenario for the red-peak GCs in this particular galaxy.

\subsection{Our approach to lift the age-metallicity degeneracy}

Optical data alone do \textbf{not} allow to disentangle ages and
metallicities: Colour-to-metallicity transformations have to assume an age,
while colour-to-age transformations are only valid for one metallicity.
The degeneracy, however, can be broken by including near-infrared data that
are more sensitive to metallicity rather than to age. 

\cite{anders04b} used extensive artificial star cluster tests and showed that
observations in three passbands for GCs in dust-free E/S0s (or four passbands
for young star clusters in dusty environments), spanning as wide as possible a
wavelength-basis ($U$ through $K$) and including at least one NIR-band (e.g.
$H$ or $K$) with photometric accuracies $\leq 0.05$ mag in the optical and
$\leq 0.1$ mag in the NIR allow to disentangle ages and metallicities and to
determine \textbf{individual GC metallicities to $< \pm 0.2$ dex, and ages to
  $< \pm 0.3$ dex}. I.e., these data allow to distinguish $\leq 7\,\rm Gyr$
old GCs from those $\geq 13\,\rm Gyr$ old.

Similar studies also using NIR-data to determine ages and metallicities of
globular clusters have only been done for a few galaxies until now
\citep{puzia02,kisslerpatig02,hempel03,larsen05,hempel07}. More than half of
these galaxies were found to host a population of GCs that is younger and/or
more metal-rich than the ubiquitous old and metal-poor GC population.

Some of these previous studies discussed average properties of the blue and
red peak GCs by investigating the mean colours of blue and red GC. We here
derive ages, metallicities, and masses for every individual cluster with
($g,~z,~K$) photometry. This also enables us to study whether there is more
than one generation of GCs in the red peak, to look for correlations of these
parameters, and investigate their spatial distributions.

We have chosen the Virgo cluster S0 galaxy NGC4570 ($=$ VCC1692) for our pilot
study because it has archival $g-$ and $z-$band data from the ACS Virgo
Cluster Project \citep{peng06} that show two very clear peaks in the optical
$g-z$ colour distribution of its GC system. The red peak is about two thirds
the height of the blue one, both have very similar widths. In particular, the
optical colour distribution shows no evidence for substructure within the red
peak or for a third peak. The mean $g-z$ colours of the blue and red peaks are
$0.88 \pm 0.01$ and $1.38 \pm 0.03$, respectively, with a fraction of 39 \% of
all the 122 GCs detected belonging to the red peak. NGC4570 has $B_T=11.82$
mag \citep{cote06}, which at a kinematic distance modulus of $BDM=31.16$
\citep{mei07} gives it an absolute $M_B = -19.3$ mag, i.e. it is an average
luminosity S0. 

\cite{vandenbosch98a} and \cite{vandenbosch98b} detected a nuclear stellar
disk with a radial extent of $\approx 7$ arcsec. From both spectroscopy based
on H$\beta$ and [MgFe] line indices as well as photometry in $U$, $V$ and $I$
they estimate an age of the stellar population inside that structure of $\leq
2$ Gyr and a metallicity close to solar.

From the same ACS Virgo Cluster survey that reported the GC colours
\cite{ferrarese06} show that NGC 4570 shows a nested-disk structure composed
of two morphologically distinct inner and out disks. Detailed isophotal
analysis reveals a blue stellar ring with radius 150 pc, but less than 7 pc
wide, around the nucleus. This ring leaves a clear imprint on the major axis
$g-$band surface brightness profile \citep[cf. Fig. 104 of][]{ferrarese06}. The
structure of this inner region was discussed by \cite{vandenbosch98b}
in the context of secular bar evolution.

\section{Models}
We used our GALEV evolutionary synthesis models for star clusters
\citep{schulz02,anders03} to compute a large grid of models for five different
metallicities $\rm -1.7 \leq [Fe/H] \leq +0.4$ and ages between $\rm 4\,Myr$
and $\rm16\, Gyr$ in time-steps of $\rm 4\, Myr$.  Our models are based on
Padova isochrones and a \cite{salpeter55} initial mass function (IMF) with a
lower-mass limit of $\rm M_{low}=0.10~M_{\sun}$ and upper mass limits between
$\rm M_{up}\approx 50-70 ~M_{\sun}$ depending on metallicity.

Since early-type galaxies in general, and NGC 4570 in particular, do not
contain significant amounts of dust we did not include internal extinction
into our grid but assume $\rm E(B-V)=0$ throughout. We therefore only need
three filters (HST F475W, F850LP and SOFI $K_s$) to determine all relevant
parameters (age, metallicity, and mass) for each cluster.

Note that we do not depend on color transformations from the HST to Standard
Johnson filters. For optimal accuracies, our models first compute spectra as a
function of time, that later are convolved with the corresponding filter
curves, in this case for the HST F475W and F850LP filters and the SOFI $K_S$
filter, to yield the final magnitudes.

\begin{figure}
\includegraphics[width=\linewidth]{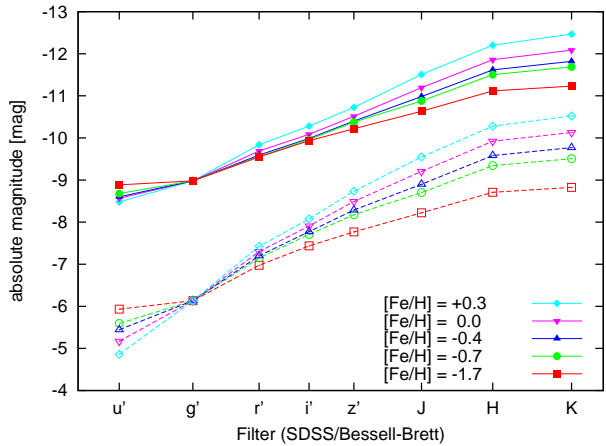}
\caption{Spectral energy distributions (SEDs) for star cluster models at two
  different ages of $\rm 1\,Gyr$ at brighter and $\rm 13\,Gyr$ at fainter
  absolute magnitudes, respectively, and 5 different metallicities ranging
  from $1/50\, Z_{\odot}$ to $2.5\,Z_{\odot}$ in each case. All SEDs have been
  scaled to match the g-band magnitude of the solar metallicity model with
  initial mass ${\rm 10^6~M_{\odot}}$.}
\label{fig:sedssdss}
\end{figure}

Figure \ref{fig:sedssdss} shows examples of Spectral Energy Distributions
(SEDs) for star cluster models at two different ages of 1 and 13 Gyr, and all
five metallicities for each of them. The SEDs shown have been scaled
  to match the g-band magnitude of a solar metallicity star clusters with
  initial mass ${\rm 10^6~M_{\odot}}$. Fluxes in all filter bands of course
  scale with cluster mass. At both ages the lines representing the different
  metallicities split up towards both longer and shorter wavelengths. We
  exploit this to separate the effects of age and metallicity.


\subsection{AnalySED}
To obtain accurate ages, metallicities and masses for the detected clusters we
used the SED analysis tool AnalySED developed by \cite{anders04b}. This
program compares the observed SED of a star cluster with a large grid of model
SEDs and finds the best-fitting match to the observations. In this
  process ages and metallicities for each cluster are derived from the
  observed spectral energy distributions. Once those parameters are found
  the observed brightnesses are translated into masses.

Since we keep the full $\chi^2$ distribution and not only the best fitting
solution we are able to determine the $1\sigma$ uncertainties for all our
derived parameters by summing up the normalized probabilities sorted from the
highest to the lowest values until an integrated probability of 0.68 is
reached. The uncertainty ranges for the different parameters are then given by
their extreme values reached within this $1\sigma$ probability range (cf.
\cite{anders04b} for details of the method and e.g. \cite{anders04a} for an
earlier application).

\section{Observations and data reduction}
\subsection{Near-infrared data}
We observed NGC 4570 in two subsequent nights (2nd to 4th March 2007) using
the ESO-NTT equipped with the SOFI near-infrared imager. This instrument
consists of a Hawaii HgCdTe 1K$\times$1K chip and $0.288''$ pixels yielding a
field-of-view of $\approx 5\times 5$ arcmin. To avoid non-linearities of the
detector we chose an detector integration time (DIT) of 6 seconds. 12 of these
exposures were internally averaged by the readout electronics, resulting in an
exposure time per frame of 72 seconds. Since NGC 4570 has an extent of only
$4' \times 1.1'$, we could use one half of the chip for the object while
obtaining a simultaneous sky-exposure in the other half of the chip, swapping
sides after each exposure and applying small shifts to avoid contamination by
bad pixels. 

\subsubsection{Reduction}
Data reduction largely followed the procedures outlined in the SOFI instrument
handbook. The individual frames were corrected for the inter-row cross-talk
and then corrected with a flat-field to remove pixel-to-pixel sensitivity
variations. These flat-fields consisted of three different components:
dome-flats corrected for their characteristic on-off pattern using the recipe
provided on the NTT-SOFI web-site; illumination correction surface obtained
from repeated observations of a standard star at different positions of the
detector to remove the illumination gradient introduced by the dome flats; a
``super-flat'' created from the sky frames of both nights to remove remaining
inhomogeneities and cosmetics from dust on the filter.  Since the observations
were carried out under non-photometric conditions we used the data from the
instruments web-page\footnote{http://www.ls.eso.org/lasilla/sciops/ntt/sofi/}
to obtain the illumination correction.

After flat-fielding all the frames we estimated the background in all frames
by iteratively clipping values larger then $3\sigma$ above the mean of the
full frame. We then used six sky-frames that were obtained closest in time,
scaled each of them individually to the sky-value of the object frame and
subtracted their average from the object frame. We varied the number of frames
to average, but found that six is the best compromise between signal-to-noise
and artifacts introduced by the pupil-rotation of the alt-az mounted
telescope.

All the sky-subtracted frames were aligned by matching coordinates of several
background galaxies in each frame to coordinates in a reference frame; we used
galaxies instead of stars because due to the high galactic longitude there
were too few stars to allow for proper matching. In a final step we stacked
all but those frames with a sky-value deviating more than 2$\sigma$ from the
average.

Combining all those exposures from both nights results a total exposure time
for our K-band image of $\rm t_{exp} \approx 25~ks \approx 9.5~hours$.

\subsubsection{Photometric calibration}
We could not base our photometric calibration on standard stars, because
during both nights observations were hampered by a varying degree of
cloudiness. We therefore compared brightness profiles of the host galaxies
with calibrated data from the 2MASS survey (see \cite{skrutskie06} 
for a review) and own observations obtained later with the SIRIUS instrument
at the InfraRed Survey Facility at the South African Astronomical
Observatory. Both results showed excellent agreement within the error
ranges of $\Delta m_K \approx 0.03$ mag, that can easily be explained by
minor differences in the filter transmission curves.  However, we account for
this uncertainty by adding the calibration error in quadrature to the
photometric errors. Calibration of exposures from both 2MASS and SIRIUS
finally relied on standard stars from the \cite{persson98} catalog, so our
K-band magnitudes are in the VEGA magnitude system.

This allows us to detect point sources down to $m_{\rm Ks}\approx 21$ mag at
the $10\sigma$-level, making the depth of our observations comparable to deep
surveys obtained with the same configuration, e.g. the K20 survey
\citep{cimatti02}.

\subsection{HST data}
The HST data consisted of two datasets taken with the Advanced Camera for
Survey (ACS) on-board the Hubble space telescope as part of the ACS
Virgo cluster survey \citep{cote+04}. For both datasets with filters F475W
($\approx$ SDSS g) and F850LP ($\approx$ SDSS z) we relied on the On-the-fly
reduction performed automatically upon retrieval from the MAST
Archive\footnote{http://archive.stsci.edu/} and using the best reference
files. We then performed an additional alignment step to ensure a match of
coordinates in both frames as good as possible.

Calibration of the HST data was done using the appropriate header entries from
the fits-files. For details on this process see the HST Data Handbook
\citep{pavlovsky05}. To avoid any unnecessary conversion between different
magnitude systems, we performed photometry in both HST filters using ST
magnitudes.

\subsection{Cluster selection}
Globular Cluster candidates were selected from the HST images using SExtractor
\citep{bertin96}. A valid detection is characterized by at least 4 adjacent
pixels with intensities of 3$\sigma$ above the local background, resulting in
two catalogs with $>1000$ objects each. We then cross-correlated these
catalogs to remove remaining spurious detections as e.g. remaining
cosmics. 

For the $\approx330$ remaining candidates we derived intrinsic source sizes
using the ISHAPE-package from BAOLAB \citep{larsen99}. This algorithm in our
case convolves a King profile with a fixed concentration parameter
$c=\frac{r_t}{r_c}=30$ (or equivalently
  $\log(\frac{r_t}{r_c})\approx1.5$) but variable radii with the instrumental
point spread function (PSF) created by TinyTim \citep{tinytim} and determines
the best fitting radius via $\chi^2$-minimization. We rejected all objects
appearing stellar-like (intrinsic radius $\rm r_i<0.2~px \approx 0.8~pc$ at a
distance $\rm D\approx 17~Mpc$ \citep{tonry01}, 17 objects) or too extended
($\rm r_i > 5~px \approx 20~pc$, 37 objects) to be a globular cluster.

\subsection{Photometry}
For the remaining $\approx280$ GC candidates we obtained aperture photometry
using commands from the ESO-MIDAS package. For the HST images we used an
aperture radius of 15 pixels and a sky annulus from 17 to 20 pixels.
To compensate for the low-intensity extended wings of the ACS PSF we
  applied an aperture correction of $0.067$ ($0.082$) mag to the F475W
  (F850LP) magnitudes according to the tabulated Enclosed Energy values of
  $0.940$ ($0.927$) determined by \cite{Sirianni+05}. Although these factors
  still depend both on the position of the GC on the detector and intrinsic
  colour of the GC, the implied uncertainties are much smaller then the
  photometric uncertainties and can therefore be neglected.

To derive photometry
for the ground-based $K_s$-imaging we transformed the coordinates from the
combined HST catalog into physical coordinates in the K-image and then used
these coordinates as center positions for the photometry. Using SExtractor to
also obtain a K-band detection catalog yielded less reliable results and
missed many of the sources found with the superior resolution of the HST. We
used an aperture of $2''$ and an aperture correction derived from stellar
photometry within the field-of-view of $\rm (0.2 \pm 0.03)~mag$. For 117
candidates we could not derive a K-band magnitude, mostly because they were
not included within or too near to the edge of the K-band field-of-view, and
in some cases because they were out-shined by the bright galaxy background.

\begin{figure}
\includegraphics[width=\linewidth]{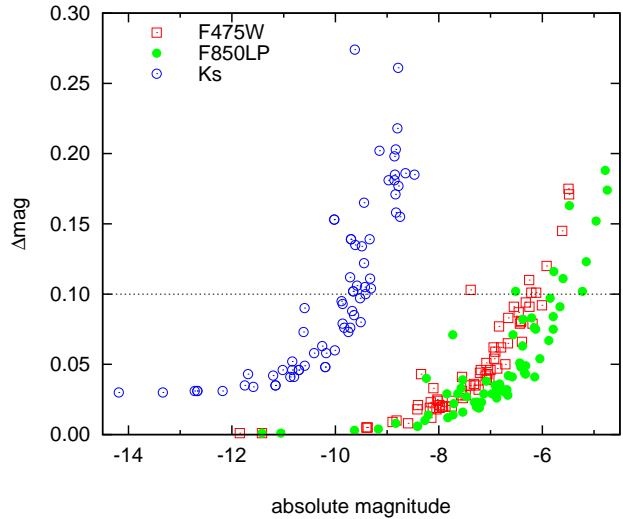}
\caption{Photometric errors as function of absolute magnitudes for the HST filters F475W and F850LP and the NTT filter $K_s$.}
\label{fig:magerror}
\end{figure}

To remove last outliers we introduced a colour selection criterion of $\rm
(g_{ST} - z_{ST}) \leq 1.0$ and $\rm (z_{ST} - K_{s,Vega}) \leq 4.5$, covering
the color range of our models. We further removed all candidates with
magnitude errors $\rm\geq 0.1~mag$ in $g$ and $z$ and $\rm\geq0.3~mag$ in $K_s$
(see Fig. \ref{fig:magerror} for the distributions of magnitude uncertainties
as a function of luminosity in all three bands).  This leaves us a final
sample of 63 {\em bona fide} GCs.

\section{Results}

\begin{figure}
\includegraphics[width=\linewidth]{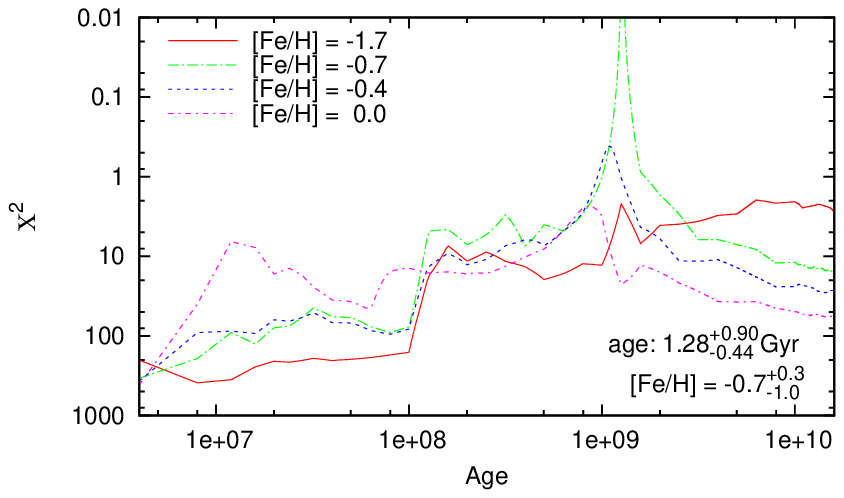}
\includegraphics[width=\linewidth]{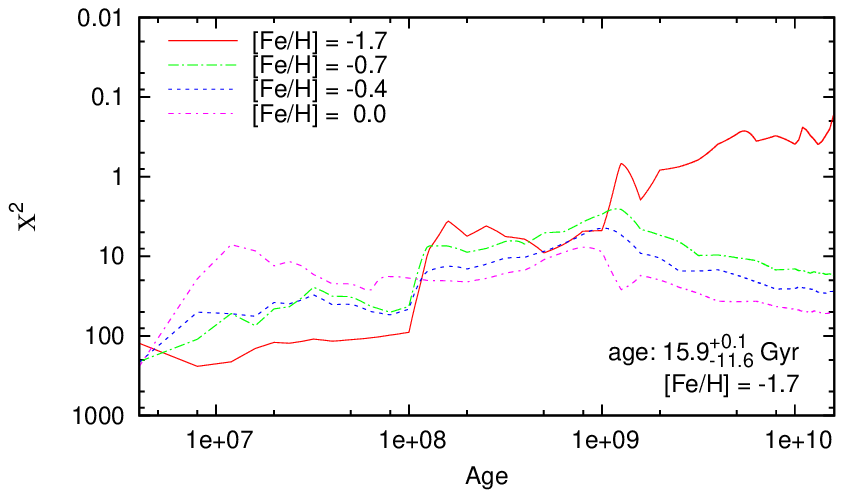}
\includegraphics[width=\linewidth]{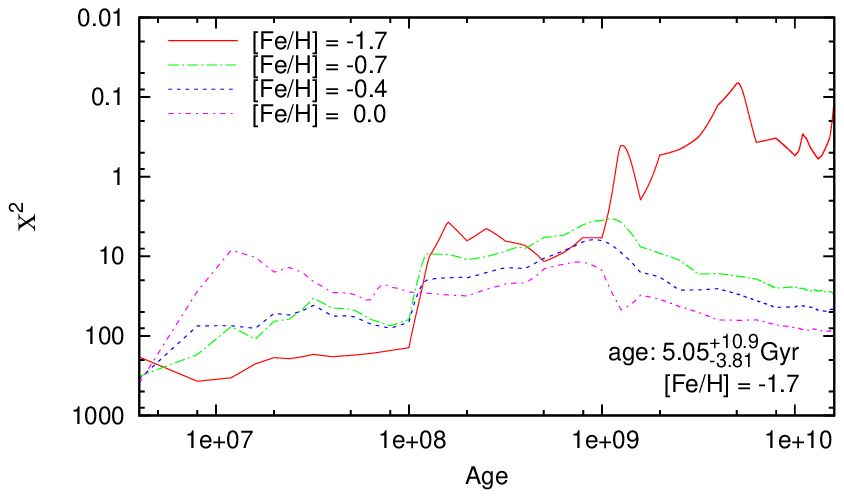}
\caption{$\chi^2$ values as function of metallicity and age for 3 different
  clusters.}
\label{fig:chi2}
\end{figure}

For three selected clusters we present the detailed $\chi^2$ distributions in
Fig. \ref{fig:chi2}. This Fig. demonstrates that the reason for the
uncertainties in the derived parameters are mostly relatively to isolated
secondary peaks in the probability distribution. The cluster in the upper
panel has a well-defined best-fit age of $1.12^{+0.90}_{-0.33}$ Gyr, a
metallicity of $\rm [Fe/H] = -0.3^{+0.3}_{-0.4}$ and a derived mass of $\rm
(2.64^{+1.59}_{-0.25})\times10^5 ~M_{\odot}$. Its $\chi^2$ value for the best
solution with ${\rm [Fe/H]=-0.4}$ is a factor of 10 better than that for the
somewhat lower metallicity ${\rm [Fe/H]=-0.7}$ paired with slightly higher
age. The cluster in the middle panel has an age of $16.0^{+0.00}_{-10.9}$ Gyr,
a metallicity of $\rm [Fe/H] = -1.7$ and a derived mass of $\rm
(4.81^{+0.00}_{-2.87})\times10^5 ~M_{\odot}$. It is clearly seen that the old
age low metallicity solution has a $\chi^2$-value about two orders of
magnitude better than that for any other metallicity. Also the intermediate
age solution for the same metallicity has a significantly higher $\chi^2$.
The cluster in the lower panel has an age of $5.62^{+10.4}_{-4.39}$ Gyr, a
metallicity of $\rm [Fe/H] = -1.7$ and a derived mass of $\rm
(9.96^{+13.6}_{-7.33})\times10^5 ~M_{\odot}$. Note the large age uncertainty
for this cluster, that directly influences the derived mass of the cluster.
Again, however, only the lowest metallicity gives an acceptable fit.
Note that since our grid of models does not sample the covered
  metallicity range finely enough. Therefore only the lowest metallicity model
  results in a good fit hence we do not give errors on metallicity for those
  two clusters.

These three GCs represent typical cases for a very good fit with a fairly well
constrained age, a securely old ($ > 5$ Gyr) but less well constrained age,
and a poorly constrained age, respectively. 

\subsection{Ages}
\begin{figure}
\includegraphics[width=\linewidth]{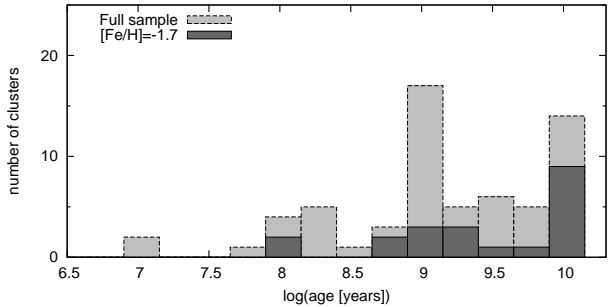}
\caption{Age distribution of our detected globular clusters.}
\label{hist:ages}
\end{figure}

Figure \ref{hist:ages} shows the age distribution of all the GCs from our final
sample. At old ages ($\rm age > 10^{10}~yr$) we see the analogues to our old
Milky Way GCs formed during an early phase of galaxy formation.

We also find a significant population of younger GCs with ages of $1-3\,\rm
Gyr$ that must have formed during some violent star formation event. Possible
candidates for such events are massive bursts of star and star cluster
formation that come along with a merger of two gas-rich spirals or the
accretion of a gas-rich companion.

While the light grey histogram in Fig. \ref{hist:ages} shows the age
distribution of all our GCs, the dark histogram shows metal-poor GCs only. The
old peak contains GCs with [Fe/H]$=-1.7$ close to the metallicity of the halo
GCs in our Milky Way, but also more metal-rich GCs while the intermediate-age
peak is predominantly made up by GCs with metallicities higher than
[Fe/H]$=-1.7$.

A detailed look at the 8 GCs in the log(age) bins 8.65-8.9 and
  9.15-9.4 shows that within their error bars they might as well be part of
  the 8.9-9.15 Gyr peak. The GCs in the age bins 9.4-9.9, on the other hand,
  could, within their $1\sigma$ uncertainties as well be part of the truely
  old, i.e. $10-13$ Gyr population.

\subsubsection{Old and metal-rich GCs}
4 out of the 10 old clusters have SEDs best described by solar metallicity.
This, however, does not seem very plausible, since it would require an
incredibly fast global chemical enrichment to solar metallicity within a
timescale of only about a Gyr. On nearer inspection, one of these
  four GCs has a large age uncertainty ranging down to 1.25 Gyrs, so it might
  well belong to our newly detected intermediate age population. The three
  remaining GCs are located within the extent of the host galaxy, so that
  their photometry might still be contaminated by light from the surrounding
  galaxy. This is also supported by the poor quality of their SED-fit
  with remarkably low unnormalized integrated probabilities $\la 10^{-10}$.

\subsection{Spatial distribution}

\begin{figure*}
\includegraphics[width=\linewidth]{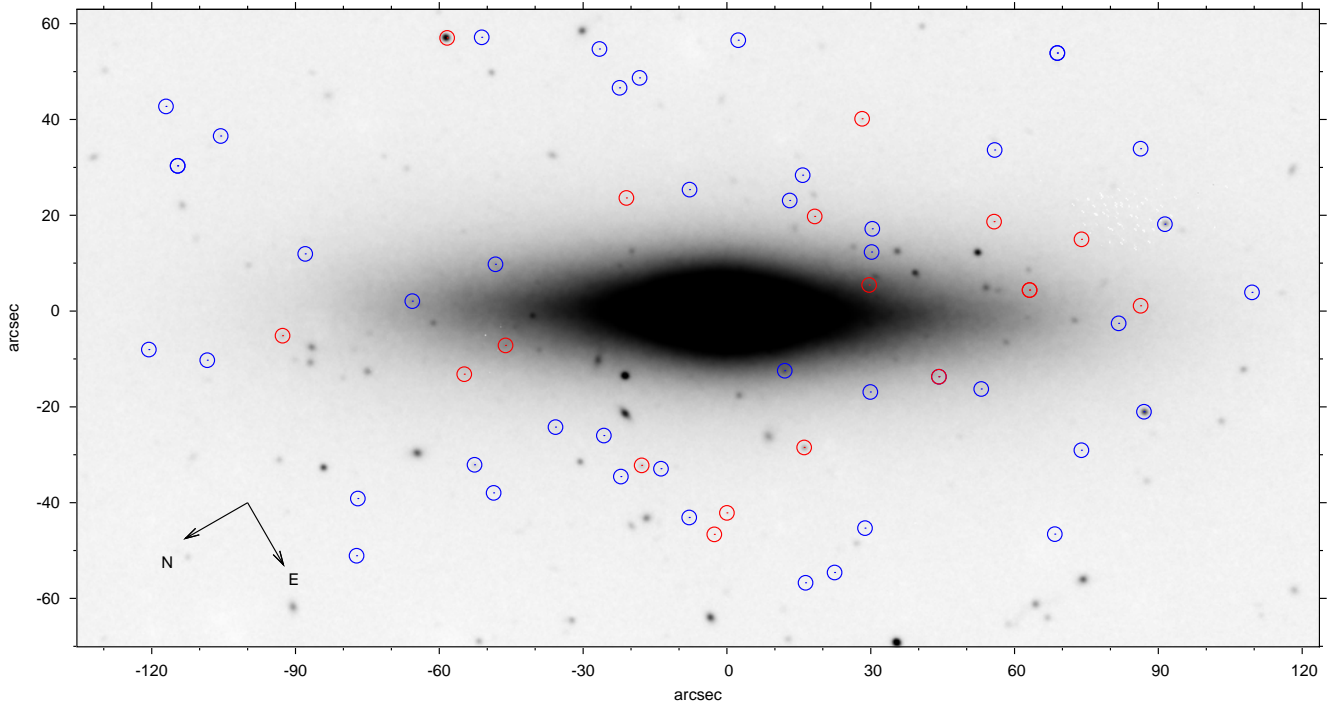}
\caption{Spatial distribution of our Globular Cluster sample,
    overlayed over our K-band image obtained with NTT/SOFI. Blue circles mark
  GCs with best-fit ages $<3$ Gyr, while red circles mark GCs older than $3$
  Gyr. Most of the bright sources are either stars or (in most cases)
  background galaxies, that are resolved on the HST images.}
\label{gc_positions}
\end{figure*}

There is no obvious difference in the spatial distributions of the old and
younger GCs, as seen in Fig. \ref{gc_positions}. GC numbers, however, are too
small for meaningful statistical tests.

We note that the bulk of the GCs we detect belong to the red peak of the
optical GC colour distribution. Only the very brightest GCs from the blue
optical peak are detected in $K_s$ and, hence, part of our sample.

\subsection{Metallicities}
\begin{figure}
\includegraphics[width=\linewidth]{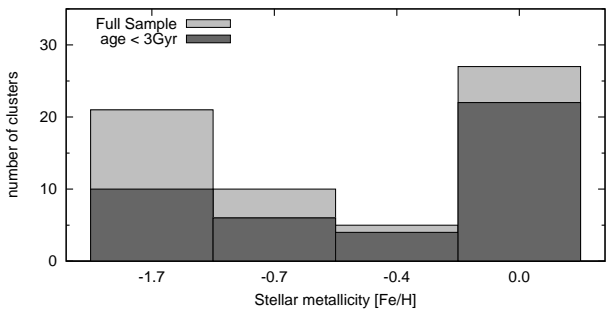}
\caption{Metallicity distribution of our GCs. Light grey histogram: all GCs, dark: GC with ages younger than 5 Gyr.}
\label{hist:metall}
\end{figure}
The metallicity distribution (Fig. \ref{hist:metall}) is dominated by
  a large population of 27 out of 63 clusters with solar metallicity, only 21
  clusters have very low metallicities of $\rm [Fe/H]=-1.7$ in the range
  of Galactic old GCs.

The fact that we find only relatively few old and metal-poor GCs can be
explained as a selection effect: Since we only include clusters that were also
detected and also have good photometry in $K_s$ we prefer intrinsically red
objects. Old Milky Way-type globulars, however, have blue colors due to their
low metallicity and are hence not included in our sample. Indeed, the bulk of
all our GCs, and of the metal-rich ones in particular, are younger than 3 Gyr,
again a selection effect, since 3 Gyr old GCs are brighter than 13 Gyr old
ones by 1.1 to 1.3 mag, depending on metallicity and filter.

\subsection{Masses}
\begin{figure}
\includegraphics[width=\linewidth]{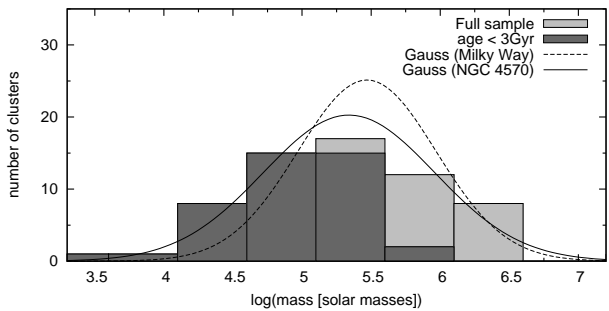}
\caption{Mass distribution of the GCs in NGC 4570. The light grey histogram is
  for all GCs from our sample, the dark one for the subsample of GCs younger
  than 3 Gyr. The solid line is a Gaussian fit to our unbinned GC mass
  distribution with a median of $\rm\log(\langle M_{GC}[M_{\odot}]
  \rangle)=5.3$ and ${\rm \sigma(M_{GC}) = 0.6~dex}$, normalized to the
  number of GCs in our sample. The dotted line gives the Gaussian for the
  Milky Way GCs with $\rm\log(\langle M_{GC}[M_{\odot}] \rangle_{MW})=5.47$
  and ${\rm \sigma_{MW}(M_{GC}) = 0.50~dex}$, normalized to the number of GCs
  in our sample.}
\label{hist:masses}
\end{figure}

Fig. \ref{hist:masses} shows the mass distribution of our GCs, again for the
full sample in light grey and for the subsample of GCs with ages $<3$ Gyr in
dark grey. The mass distribution of the full GC sample clearly looks like a
Gaussian with a turn-over mass around $\rm\log(\langle M_{GC}[M_{\odot}]
\rangle)=5.3$ and ${\rm \sigma(M_{GC}) = 0.6~dex}$, very similar to the
turn-over of the GC mass function in the Milky Way which occurs at
$\rm\log(\langle M_{GC}[M_{\odot}] \rangle_{MW})=5.47$ with
$\rm\sigma_{MW}(M_{GC}) = 0.50~dex$ \citep{ashman95}. The fact that the mass
distribution of the young GCs does not extend to the same high masses as that
for the old ones cannot be a selection effect, since more massive clusters of
young age would be easily detectable if they were there.  It seems that the
secondary event that formed the $\sim 2$ Gyr old GCs in NGC 4570 did not form
the GCs with the same mass function as the old GC population. Careful modeling
of cluster destruction effects were required to really prove this conjecture.
Dynamical friction, which is most important for massive clusters, might have
preferentially destroyed those, in particular if the secondary GC population
were more centrally concentrated than the primary one, as previously found in
many, but not all, bimodal GC systems. In addition to that we might have
missed massive GCs in front of the bright galaxy background if they either
were initially closer to the centre (mass segregation) or driven towards the
centre by dynamical friction faster than the lower mass ones.


\subsection{The subsample of GCs with high photometric accuracy}

\begin{figure}
\includegraphics[width=\linewidth]{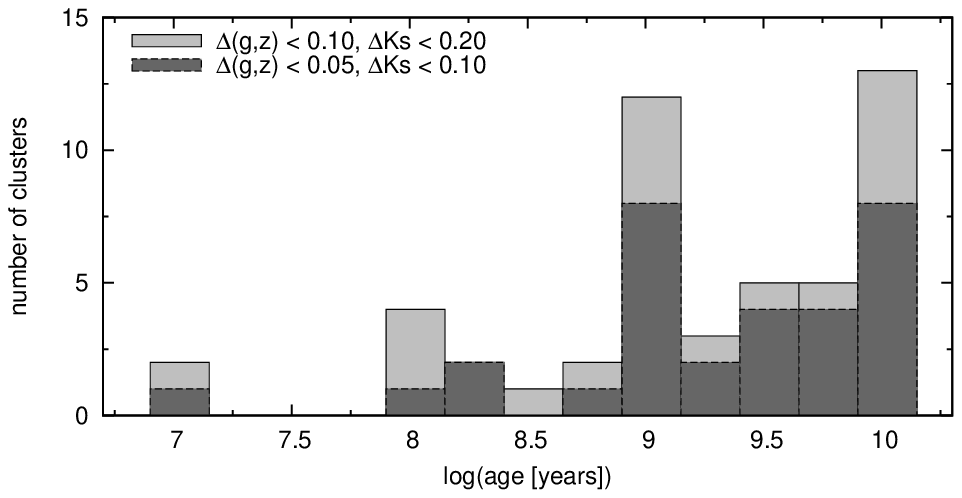}
\includegraphics[width=\linewidth]{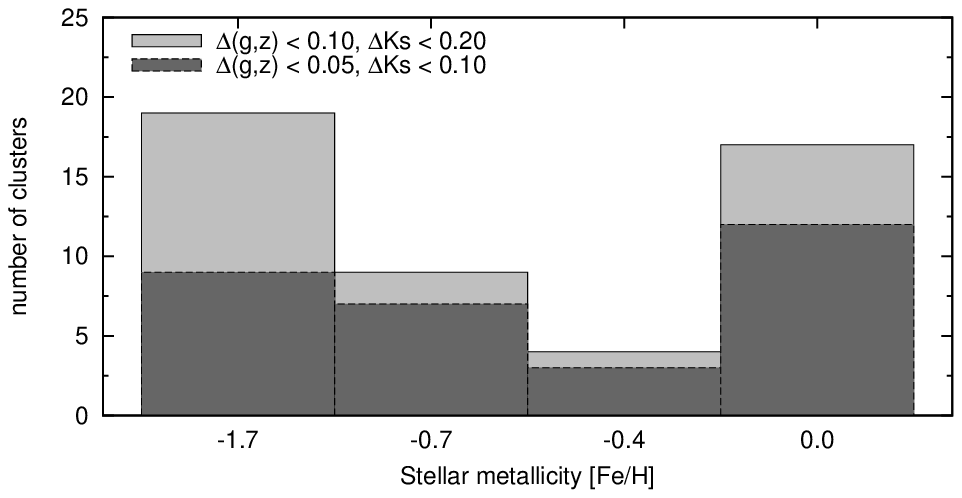}
\includegraphics[width=\linewidth]{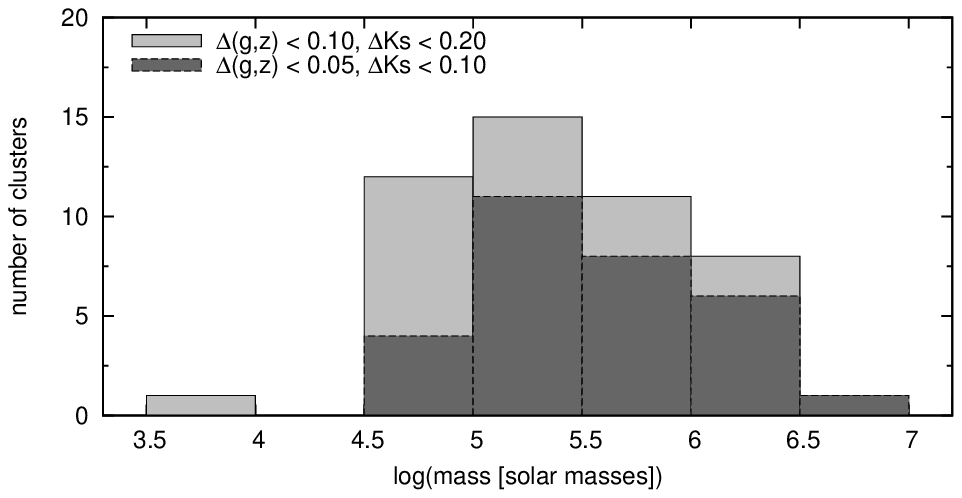}
\caption{Age, metallicity, and mass distributions for the subsample of GCs with high (light grey) and very high (dark grey) photometric accuracies.}
\label{hist:bestphot}
\end{figure}

In Fig. \ref{hist:bestphot} we present our results for the age, metallicity,
and mass distributions of the subset of GCs from our sample with the best
photometric accuracies: ${\rm \Delta g, \Delta z <0.10,~\Delta K_s <
  0.20~mag}$ and ${\rm \Delta g, \Delta z <0.05,~\Delta K_s < 0.10~mag}$,
respectively. This Fig. shows that our results are unaffected by the slightly
higher uncertainties in our full sample. The age, metallicity, and mass
distributions are robust. High photometric accuracies are only achieved for
the brightest clusters.  Hence it is no surprise that the high accuracy
subsample lacks some of the low mass GCs.

\subsection{Alternative solutions}


\subsubsection{Ages for fixed metallicities $\rm [Fe/H]=-1.7$ or $\rm [Fe/H]=0.0$}

In this subsection we explore to what extent our results change if we
artificially restrict the allowed parameter range. Our aim is to see whether
our basic finding of $\sim 1-3$ Gyr old/young GCs in NGC 4570 is affected by
these assumptions.
 
\begin{figure}
\includegraphics[width=\linewidth]{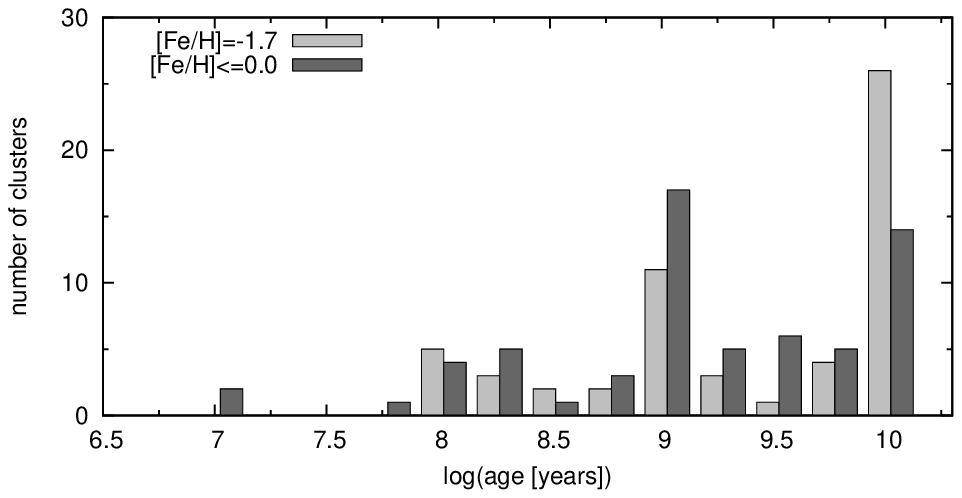}
\includegraphics[width=\linewidth]{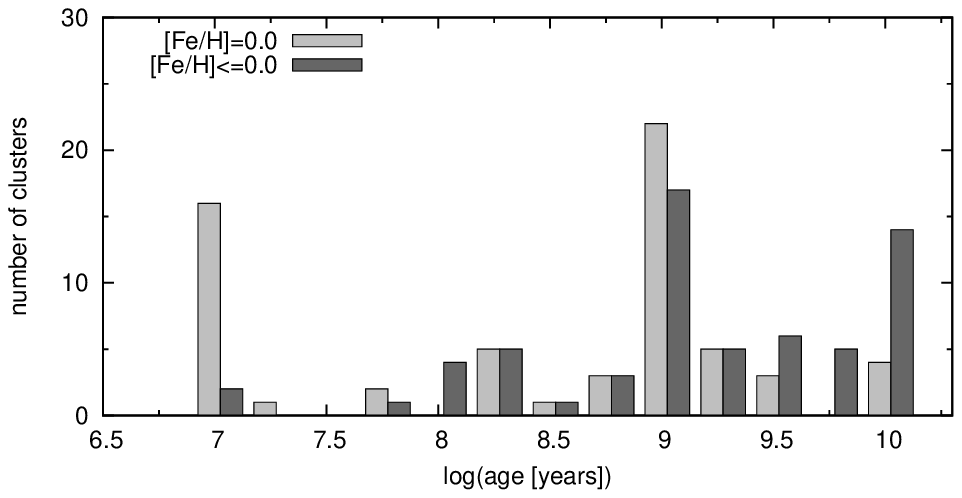}
\caption{Age distributions of our GC sample under the assumption of a fixed
  low metallicity $\rm [Fe/H]=-1.7$ (upper panel) and $\rm [Fe/H]=0.0$ (lower
  panel).  Bin sizes are the same as in Fig \ref{hist:ages}, only the two
  histograms for free and restricted metallicities are plotted next to each
  other to better reveal the differences.}
\label{hist:ages_z0}
\end{figure}
If we limit the available parameter range to a metallicity $\rm [Fe/H]=-1.7$
(see. upper panel of Fig \ref{hist:ages_z0} for comparison) many clusters are
better described by older ages. These are needed to counter-balance the bluer
colours of the lower metallicity models. The older ages, in turn, also result
in slightly higher masses.  However, the principle structure of the age
distribution and, in particular, the peak at intermediate ages of 1--3 Gyr
remain. This ``worst-case'' scenario emphasizes the robustness of our cluster
age determination.

We also determined ages for a metallicity fixed to the solar value and again
found little difference compared to the original results (see lower panel of
Fig. \ref{hist:ages_z0}). Although clusters on average get slightly younger in
this case, the peak of the distribution around 1--3 Gyr still remains, only
the tail towards younger ages become slightly more pronounced.


We conclude, that our finding of a substantial population of intermediate-age
GCs with ages of order 1 -- 3 Gyr and approximately solar metallicities in NGC
4570 is very robust, even under extreme test changes in metallicity.

\section{Discussion}
\subsection{Formation scenarios for red-peak GCs}
Our detection of a substantial population of 1 -- 3 Gyr old GCs is most
naturally explained in terms of a merger or accretion event which involved
significant amounts of gas and triggered a strong starburst in which, together
with some hitherto undetected population of field stars, the presently
observed GC formed. The star clusters we find have all the properties of GCs,
they are compact with half-light radii of order 3--7 pc and have masses in the
range of typical GC masses. Those clusters that we observe with ages of 1 -- 3
Gyr certainly are only a small fraction of all the star clusters formed in
such an event. They have already survived the most dangerous phase in their
lives, the infant mortality and dynamical restructuring phase after the first
SNe have expelled the gas left-over from their formation. They probably have
also survived the violent relaxation phase that restructured the
merging/accreted galaxies into the presently observed S0. The sheer number of
young GCs requires substantial gas masses involved and their high
metallicities indicate that it could not have been a minor accretion event
involving e.g. an SMC type galaxy swallowed by some major gas-free/poor
E/S0/Sa type galaxy. Solar metallicities in the ISM out of which the new GCs
were formed require at least one Sbc- or Sb-type progenitor. Sa-type galaxies
have higher metallicities, but lack the required amounts of gas to fuel the
starburst. 

It is intriguing that the age we find for the young GCs agrees very well with
the stellar population age of $\leq \rm 2$ Gyr estimated by
\cite{vandenbosch98b} for the nuclear stellar disk in the host galaxy
\citep{vandenbosch98a,vandenbosch98b,scorza98}.  Our detection of a
substantial population of GCs does not, however, support the internal bar
instability scenario they favor for the formation of this nuclear stellar
disk. Nuclear stellar disks are in fact seen in dynamical simulations of
galaxy mergers/accretion events
\citep[cf.][]{Bournaud+04,SpringelHernquist05}, as well as in real merger
remnants, e.g. in Arp 214 and Arp 224 \citep{JogChitre02}. The fact that NGC
4570 does not feature any obvious tidal tails can be understood as a
consequence of its location in a high galaxy density region of the Virgo
cluster, where tidal tails are shred down as soon as they start to develop.

We hence suggest that the young GCs we detected and the nuclear stellar disk
in NGC 4570 have been formed by the same merger/accretion event.

\section{Future prospects}
HST NIC3 data would be very valuable in order to detect young GC closer to the
centre of NGC 4570 and see whether the young GC system is more centrally
concentrated than the old one or not. Of equal value would be an
  extension of the SED to shorter wavelengths. As can also be seen from Figure
  \ref{fig:sedssdss} for both old and young clusters the U-band is both
  sensitive to ages and metallicity and can help discriminate between both.
  However, imaging in further intermediate band-passes would only be necessary
  for other galaxies containing dust \citep[for further details see
  ][]{anders04b}.  Spectroscopy, albeit challenging, would allow to confirm
and eventually further constrain the photometric ages, metallicities and
masses for these young GCs, while abundance ratios, e.g. of [$\alpha$/Fe]
could give further clues to the progenitor galaxies and to the star cluster
formation scenario.

A natural next step seem to combine our results from this GC analysis with a
detailed analysis of the stellar population across the main body of the galaxy
to evaluate the contribution (and location) of field stars from the age range
of the young GCs. This could either be done on the basis of multi-band imaging
in a pixel-by-pixel analysis of the kind we did for the Tadpole and Mice
galaxies \citep{deGrijs+03} or analyzing the integrated spectrum with
starburst models as we have done for the $\sim 1$ Gyr old merger remnant NGC
7252 \citep[cf.][]{FritzeGerhard94a,FritzeGerhard94b}.

\section{Summary}
On the basis of deep NTT-SOFI $K_s-$ band imaging in conjunction with archival
HST ACS deep optical imaging we identified a substantial population of star
clusters with ages in the range 1 -- 3 Gyr and metallicities around solar in
the Virgo S0 galaxy NGC 4570. All these clusters are compact with half-light
radii in the range 3--7 pc and have masses of order ${\rm 10^5~M_{\odot}}$.
They have successfully survived their infant mortality phase and, hence, merit
to be called young GCs. The clusters we detect in $K_s$ make up an important
fraction of the red peak of the bimodal GC optical colour distribution
reported by \cite{peng06} from the ACS Virgo Cluster Survey.

We performed a number of test that showed our results to be robust. 

The ages we find for the young GCs agree well with stellar population ages
previously determined for the nuclear stellar disk in NGC 4570. We suggest
that both the nuclear stellar disk and the young GC population have been
formed in the same merger/accretion event. The presently observed GCs
certainly are only a small part of the originally formed star cluster
population, since only a small fraction of clusters usually survives until
ages of $\geq 1$ Gyr. The sheer number of young GCs suggests that the
merger/accretion event must have involved substantial gas masses, the high
metallicity of the new GCs suggests that the gas involved in this event must
have been enriched at least to a level observed in present-day Sbc galaxies.

Our analysis has shown that GC populations are valuable tracers of their
parent galaxy's star formation, chemical enrichment, and mass assembly
histories.

Clearly further analyses, in particular of the stellar population across the
main body of NGC 4570, are required before we fully understand the details of
the scenario that gave rise to the present-day Virgo S0 NGC 4570 with its
nuclear stellar disk and young GCs.

\section*{Acknowledgments}
We thank the International Space Science Institute (ISSI) for their
hospitality and research support and our anonymous referee for very
  insightful comments that helped improve and clarify this paper. We are also
grateful to Hagen Meyer who obtained the calibration data at the IRSF. RK
thanks the ESO-NTT crew, Valentin Ivanov, Alessandro Ederoclite, and Monica
Castillo, for support during the preparation and execution of the
observations.

This publication is based on observations made with ESO Telescopes at the La
Silla Observatory under programme ID 079.B-0511. This paper is also based on
archival observations with the NASA/ESA Hubble Space Telescope, obtained at
the Space Telescope Science Institute, which is operated by the Association of
Universities for Research in Astronomy (AURA), Inc, under NASA contract NAS
5-26555.

\bibliographystyle{mn2e} 
\bibliography{gcs_ngc4570}
\label{lastpage}

\end{document}